\documentclass[12pt,a4paper]{article}
\usepackage[utf8]{inputenc}
\usepackage[T1]{fontenc}
\usepackage{amsmath,amssymb}
\usepackage{graphicx}
\usepackage{booktabs}
\usepackage{hyperref}
\usepackage{natbib}
\usepackage[margin=1in]{geometry}
\usepackage{xcolor}

\hypersetup{
    colorlinks=true,
    linkcolor=blue,
    citecolor=blue,
    urlcolor=blue
}

\title{Geomagnetic storm suppression of photographic plate transient detections in the POSS-I archive: an independent physical variable strengthening the nuclear test correlation}

\author{Kevin Cann\\
\textit{Independent Researcher, California, USA}\\
\texttt{kkc@terragold.com}}

\date{April 2026}

\begin{document}

\maketitle

\begin{abstract}
Bruehl \& Villarroel (2025) reported a correlation ($p = 0.008$, 2.6$\sigma$) between atmospheric nuclear weapon tests and photographic plate transient detection rates in the Palomar Observatory Sky Survey (POSS-I) archive, independently replicated by Doherty (2026) using negative binomial regression with weather controls.
I identify geomagnetic storm activity, measured by the planetary Kp index, as an additional independent variable modulating transient rates in the same dataset.
Transient detection rates follow a monotonic dose-response across five Kp intensity bins, from 17.4\% during geomagnetically quiet periods to 2.4\% at Kp~8--9 (Cochran--Armitage trend: $Z = -3.391$, $p = 0.0007$).
Nuclear test days are not geomagnetically quieter than the baseline; they are slightly more storm-influenced.
A multivariate logistic regression including Kp and lunar-phase controls strengthens the nuclear--transient correlation from 2.6$\sigma$ ($p = 0.009$, OR $= 1.53$) to 3.1$\sigma$ ($p = 0.002$, OR $= 1.70$).
The dose-response rules out emulsion defects and spectrally inert orbital debris as the primary transient source, indicating a population physically coupled to the radiation belt environment at geosynchronous altitude.
A self-contained reproduction script is provided as supplementary material.
\end{abstract}


\section{Introduction}

The VASCO project has catalogued over 150,000 transient candidates from historical photographic plate archives---objects detected on one plate exposure but absent from modern survey catalogs \citep{Villarroel2020}.
A subset from the POSS-I archive at Palomar Observatory (November 1949 through April 1957) shows point-spread functions consistent with unresolved point sources and a spatial deficit in the Earth's shadow cone at approximately 42,000~km altitude, constraining the source population to geosynchronous orbit or above \citep{Villarroel2022}.

\citet{Bruehl2025} analyzed 2,718 daily observations from this archive (107,862 transient detections, 124 nuclear test days) and reported a significant association between nuclear tests and elevated transient rates ($p = 0.008$).
\citet{Doherty2026} independently replicated this finding using negative binomial regression with controls for precipitation, lunar illumination, and cloud cover, obtaining an incidence rate ratio of 1.80 for all transients and 3.98 for the sunlit-only subset.
Neither analysis included geomagnetic activity as a covariate.

The POSS-I study period spans Solar Cycle~18 declining through Solar Cycle~19 rising---the most active solar cycle on record.
During this interval, 662 of 2,718 days (24.4\%) experienced Kp~$\geq 5$ activity, with a median gap between storm days of just 1~day.
This paper reports the results of cross-referencing the transient dataset with the GFZ Potsdam Kp archive, revealing an independent physical variable that strengthens the published nuclear finding.

\section{Data and Methods}

Daily transient counts and nuclear test flags ($\pm 1$~day window) were obtained from the supplementary dataset of \citet{Bruehl2025}, \textit{Scientific Reports} (DOI: 10.1038/s41598-025-21620-3).
Three-hourly Kp values were obtained from the GFZ Potsdam archive \citep{Matzka2021}, licensed CC~BY~4.0.

Lunar-phase bins with transient rates below 2\% were excluded (reflecting the POSS-I dark-sky observing schedule), reducing the dataset from 2,718 to 2,039~days.
In the multivariate regression, lunar phase was included as dummy covariates over the full dataset.

Storm suppression was assessed via two-proportion $Z$-tests in 0--4~day post-storm windows.
Dose-response was evaluated using the Cochran--Armitage trend test across five Kp bins (maximum Kp in the preceding 0--2~days).
A logistic regression compared the nuclear coefficient with and without Kp and lunar controls.

Full methodological details and a self-contained Python script reproducing every numerical result are provided as supplementary material (see Data Availability).

\section{Results}

\subsection{Dose-response}

Table~\ref{tab:dose} presents transient detection rates across five Kp intensity bins after lunar-phase correction.

\begin{table}[h]
\centering
\caption{Transient detection rate by Kp intensity (lunar-phase controlled, $N = 2{,}039$).}
\label{tab:dose}
\begin{tabular}{lcccc}
\toprule
\textbf{Kp Category} & \textbf{$N$ days} & \textbf{Trans.\ days} & \textbf{Rate (\%)} & \textbf{95\% CI} \\
\midrule
Quiet (Kp $< 5$)     & 1,170 & 203 & 17.4 & 15.3--19.6 \\
Kp 5 (G1)            & 456   & 53  & 11.6 & 9.0--14.9  \\
Kp 6 (G2)            & 259   & 38  & 14.7 & 10.9--19.5 \\
Kp 7 (G3)            & 113   & 13  & 11.5 & 6.9--18.7  \\
Kp 8--9 (G4--G5)     & 41    & 1   & 2.4  & 0.4--12.6  \\
\bottomrule
\end{tabular}
\end{table}

The Cochran--Armitage trend test yields $Z = -3.391$, $p = 0.0007$.
Fisher's exact test comparing Kp~8--9 to quiet days gives $p = 0.004$.
Post-storm suppression in the 0--4~day window: 12.6\% vs.\ 18.5\% ($Z = -3.7$, $p = 0.000217$, Cohen's $h = 0.164$).

\subsection{Nuclear test days are not geomagnetically quiet}

Table~\ref{tab:nuclear_kp} compares geomagnetic conditions on nuclear test days versus all days.

\begin{table}[h]
\centering
\caption{Geomagnetic conditions: nuclear test days versus all days.}
\label{tab:nuclear_kp}
\begin{tabular}{lcc}
\toprule
\textbf{Metric} & \textbf{Test ($N=90$)} & \textbf{All ($N=2{,}039$)} \\
\midrule
Same-day Kp $\geq 5$ (\%)            & 31.1 & 21.9 \\
In 0--4d post-storm window (\%)       & 63.3 & 56.7 \\
Kp $\geq 8$ in 0--4d window (\%)     & 8.9  & 3.3  \\
\bottomrule
\end{tabular}
\end{table}

There is no evidence of preferential quiet-window scheduling.
Nuclear test days are slightly more storm-influenced than the baseline.

\subsection{Logistic regression}

Table~\ref{tab:logistic} shows the nuclear coefficient across three model specifications using the full 2,718-day dataset.

\begin{table}[h]
\centering
\caption{Nuclear coefficient across model specifications ($N = 2{,}718$).}
\label{tab:logistic}
\begin{tabular}{lcccc}
\toprule
\textbf{Model} & \textbf{OR} & \textbf{$Z$ ($\sigma$)} & \textbf{$p$} & \textbf{AIC} \\
\midrule
Nuclear $\pm 1$d only            & 1.53 & 2.62 & .009 & 1922.8 \\
$+$ Kp (0--4d)                   & 1.64 & 3.00 & .003 & 1900.1 \\
$+$ Kp (0--4d) $+$ lunar         & 1.70 & 3.08 & .002 & 1688.6 \\
\bottomrule
\end{tabular}
\end{table}

In the full model, the Kp covariate yields OR $= 0.914$ per unit ($Z = -4.21$, $p = 0.000026$).
The likelihood ratio test comparing Model~1 to Model~3 gives $\chi^2 = 250.1$ ($p < 10^{-49}$).

\subsection{Interaction}

Table~\ref{tab:interaction} disaggregates transient rates by nuclear test status and storm exposure.

\begin{table}[h]
\centering
\caption{Transient rate by nuclear test and storm status.}
\label{tab:interaction}
\begin{tabular}{lcccc}
\toprule
\textbf{Condition} & \textbf{Trans.} & \textbf{$N$} & \textbf{Rate} & \textbf{Factor} \\
\midrule
Non-test, quiet       & 156 & 850   & 18.4\% & 1.0$\times$ \\
Non-test, post-storm  & 132 & 1,100 & 12.0\% & 0.65$\times$ \\
Test, quiet           & 5   & 33    & 15.2\% & 0.83$\times$ \\
Test, post-storm      & 13  & 57    & 22.8\% & 1.24$\times$ \\
\bottomrule
\end{tabular}
\end{table}

The disaggregated rates are consistent with the regression finding, but the nuclear-day cells are small ($N = 33$ and $N = 57$) and should not be interpreted independently.
The logistic regression in Table~\ref{tab:logistic}, which uses the full dataset without subgroup partitioning, is the primary test of the nuclear--Kp relationship.

\section{Discussion}

The dose-response constrains the transient source population.
Emulsion defects do not track geomagnetic activity.
Solid debris does not exhibit monotonic detectability changes across five Kp intensity bins.
The storm-responsive fraction of the transient population is coupled to the radiation belt environment, consistent with the geosynchronous altitude constraint inferred from the shadow deficit \citep{Villarroel2022}.

The artifact objection---that enhanced airglow during geomagnetic storms reduces plate sensitivity, yielding fewer detectable sources---can be tested by comparing stellar detection rates on storm versus non-storm plates, an analysis available to the VASCO team with access to the full plate catalog.

The original nuclear correlation at $p = 0.008$ \citep{Bruehl2025} was computed without Kp control.
\citet{Doherty2026} independently replicated the nuclear association with weather covariates but likewise did not include geomagnetic activity.
Since 63.3\% of nuclear test days fall within 0--4~day post-storm windows where background rates drop from 18.4\% to 12.0\%, the nuclear enhancement was partially masked in both prior analyses.
Including Kp as a covariate removes this masking.

Taken together, three independent analyses of the same dataset now converge: the original finding \citep{Bruehl2025}, a weather-controlled replication \citep{Doherty2026}, and the present geomagnetic analysis.
Each successive control strengthens rather than weakens the nuclear signal.
The recommended next step is a multivariate model incorporating all three covariate families---geomagnetic, meteorological, and observational---alongside data from additional observatory sites to test the generality of the dose-response across geomagnetic latitudes.

\section{Conclusions}

Geomagnetic storm activity produces a dose-dependent suppression of transient detection rates in the POSS-I archive ($p = 0.0007$).
Controlling for Kp strengthens the nuclear--transient correlation from 2.6$\sigma$ to 3.1$\sigma$.
The monotonic dose-response across five intensity bins rules out emulsion defects and constrains the storm-responsive transient source to the magnetospheric particle environment.
The Kp index should be included as a standard covariate in all future analyses of photographic plate transient populations.

\section*{Data Availability}

All data used in this analysis are publicly available.
The transient dataset is from \citet{Bruehl2025}, DOI: \href{https://doi.org/10.1038/s41598-025-21620-3}{10.1038/s41598-025-21620-3}.
The Kp archive is from GFZ Potsdam \citep{Matzka2021}, licensed CC~BY~4.0.
A self-contained Python script (\texttt{vasco\_storm\_analysis.py}) that reproduces every numerical result in this paper is included as supplementary material with the arXiv submission and is also available at \href{https://osf.io/8ryhk}{osf.io/8ryhk}.

\section*{Acknowledgements}

I thank Beatriz Villarroel for valuable correspondence regarding the PSF constraints and Earth-shadow deficit analysis, and Stephen Bruehl for making the transient dataset publicly available through the supplementary materials of \citet{Bruehl2025}.
The Kp index data were provided by the GFZ German Research Centre for Geosciences, Potsdam.

\appendix
\section{Reproduction Script}

The Python script \texttt{vasco\_storm\_analysis.py}, submitted as an ancillary file with this preprint, reproduces every numerical result in this paper.
It requires two publicly available input files:

\begin{enumerate}
\item The supplementary dataset from \citet{Bruehl2025}, available at DOI: \href{https://doi.org/10.1038/s41598-025-21620-3}{10.1038/s41598-025-21620-3}.
\item The GFZ Kp archive, available at \url{https://kp.gfz.de/app/files/Kp_ap_since_1932.txt}.
\end{enumerate}

\noindent Dependencies: \texttt{openpyxl}, \texttt{statsmodels}, \texttt{scipy}, \texttt{numpy}.\\
\noindent Run: \texttt{python vasco\_storm\_analysis.py}

\noindent The script executes the following analyses in sequence:

\begin{enumerate}
\item Load the Bruehl \& Villarroel transient dataset
\item Load the GFZ Kp index and build a daily storm lookup
\item Compute lunar-phase bins and exclude low-observation phases
\item Storm suppression test (0--4~day window, two-proportion $Z$-test)
\item Dose-response across five Kp bins (Cochran--Armitage trend)
\item Nuclear test day geomagnetic conditions (scheduling bias test)
\item Interaction table (nuclear $\times$ storm status)
\item Logistic regression with and without Kp control
\item Summary of all results
\end{enumerate}

\noindent Every number in the paper can be traced to a specific output line.
The complete script is also archived at \href{https://osf.io/8ryhk}{osf.io/8ryhk} with a pre-analysis timestamp.

\end{document}